\documentclass{amsart}
\makeatletter

\newtheorem{thm}{Theorem}[section]
\def\statetheorem{\@ifnextchar[{\@statetheorem}{\nr@statetheorem}}
\long\def\@statetheorem[#1]#2{\begin{thm}\label{#1}#2\end{thm}}
\long\def\nr@statetheorem#1{\begin{thm}#1\end{thm}}
\def\statetheorempf{\@ifnextchar[{\@statetheorempf}{\nr@statetheorempf}}
\long\def\@statetheorempf[#1]#2{\begin{thm}\label{#1}#2\end{thm}\proof}
\long\def\nr@statetheorempf#1{\begin{thm}#1\end{thm}\proof}

\newtheorem{lmma}{Lemma}[section]
\def\statelemma{\@ifnextchar[{\@statelemma}{\nr@statelemma}}
\long\def\@statelemma[#1]#2{\begin{lmma}\label{#1}#2\end{lmma}}
\long\def\nr@statelemma#1{\begin{lmma}#1\end{lmma}}
\def\statelemmapf{\@ifnextchar[{\@statelemmapf}{\nr@statelemmapf}}
\long\def\@statelemmapf[#1]#2{\begin{lmma}\label{#1}#2\end{lmma}\proof}
\long\def\nr@statelemmapf#1{\begin{lmma}#1\end{lmma}\proof}

\newtheorem{crlry}{Corollary}[section]
\def\statecorollary{\@ifnextchar[{\@statecorollary}{\nr@statecorollary}}
\long\def\@statecorollary[#1]#2{\begin{crlry}\label{#1}#2\end{crlry}}
\long\def\nr@statecorollary#1{\begin{crlry}#1\end{crlry}}
\def\statecorollarypf{\@ifnextchar[{\@statecorollarypf}{\nr@statecorollarypf}}
\long\def\@statecorollarypf[#1]#2{\begin{crlry}\label{#1}#2\end{crlry}\proof}
\long\def\nr@statecorollarypf#1{\begin{crlry}#1\end{crlry}\proof}

\def\ord{\textup{ord}\,}

\def\C{\mathbb{C}}
\def\N{\mathbb{N}}
\def\bibref[#1]{\cite{#1}}

\let\real@bibitem\bibitem
\def\bibitem[#1]{\real@bibitem{#1}}

\makeatother

\title{Bispectrality of KP Solitons}

\author{Alex Kasman}
\address{Department of Mathematics and Statistics, Concordia
University \\and\\ Centre de recherches math\'ematiques, Universit\'e de
Montr\'eal}

\curraddr{Mathematical Sciences Research Institute, Berkeley,
CA}

\def\hatTC{{\hat K}_C}
\def\genT{{\hat T}}
\def\hatGamma{{\hat\Gamma}}
\def\hatLambda{{\hat\Lambda}}
\def\TC{{\hat Q}_C}
\def\D{{\mathbb{D}}}

\def\TODOs{{\mathbb{T}}}
\def\L{{\mathcal L}}
\def\Trans{{\textbf{S}}}
\begin{document}

\begin{abstract}{It is by now well known that the wave functions of rational
solutions to the KP hierarchy which can be achieved as limits of the
pure $n$-soliton solutions satisfy an eigenvalue equation for ordinary
differential operators in the spectral parameter.  This property is
known as ``bispectrality'' and has proved to be both interesting and
useful.  In this note, it is shown that certain (non-rational) soliton
solutions of the KP hierarchy satisfy an eigenvalue equation for a
non-local operator constructed by composing ordinary differential
operators in the spectral parameter with {\it translation\/} operators
in the spectral parameter, and therefore have a form of bispectrality
as well.  Considering the results relating ordinary bispectrality to
the self-duality of the rational Calogero-Moser particle system, it
seems likely that this new form of bispectrality should be related to
the duality of the Ruijsenaars system.  }\end{abstract}

\maketitle

\section{Introduction}
\subsection{The KP Hierarchy and Bispectrality}
Let $\D$ be the vector space spanned over $\C$ by the set
$$
\{\Delta(\lambda,n)\mid \lambda\in\C, n\in\N \}
$$
whose elements differentiate and evaluate functions of the variable
$z$:
$$
\Delta(\lambda,n)[f(z)]:=f^{(n)}(\lambda).
$$
The elements of $\D$ are thus \emph{finitely supported
  distributions} on appropriate spaces of functions in $z$.  For lack
of a better term, we will continue to call them distributions even
though their main use in this paper will be their application to
functions of two variables.    (Such
distributions were called ``conditions'' in \bibref[W] since a
KP wave function was specified by requiring that it be in their
kernel.)  Note that if $c\in\D$ and $f(x,z)$
is sufficiently differentiable in $z$ on the support of $c$, then
$\hat f(x)=c[f(x,z)]$ is a function of $x$ alone.  Furthermore,
note that one may ``compose'' a distribution with a function of $z$,
i.e. given $c\in\D$ and $f(z)$ (sufficiently differentiable on the
support of $c$) then there exists a $c':=c\circ f\in\D$ such
that
$$
c'(g(z))=c(f(z)g(z))\qquad\forall g.
$$

The subspaces of $\D$ can be used to generate solutions to the KP
hierarchy \bibref[SW] in the following way.  Let $C\subset\D$ be an
$n$ dimensional subspace with basis $\{c_1,\ldots,c_n\}$.  Then, if
$K=K_C$ is the unique, monic ordinary differential operator in $x$ of
order $n$ having the functions $c_i(e^{xz})$ in its kernel (see
\eqref{eqn:wrformK}) we define $\L_C=K \frac{\partial}{\partial x} K^{-1}$ and
$\psi_C=\frac{1}{z^n}K e^{xz}$.  The connection to integrable systems
comes from the fact that adding dependence to $C$ on a sequence of
variables $t_j$ ($j=1,2,\ldots$) by letting $C(t_j)$ be the space with
basis $$\{c_1\circ e^{\sum -t_jz^j},c_2\circ e^{\sum -t_jz^j},\ldots,
c_n\circ e^{\sum -t_jz^j}\}$$ it follows that the ``time dependent''
pseudo-differential operator $\L=\L(t_j)$ satisfies the equations of
the KP hierarchy \bibref[BHYsato,thesis,cmbis,Liberati,W] $$
\frac{\partial}{\partial t_j}\L=[(\L^j)_+,\L].
$$
The {\it wave function\/}
$\psi_C(x,z)$ generates the corresponding subspace of the infinite
dimensional grassmannian $Gr$ \bibref[SW] which parametrizes KP solutions and
thus it is not difficult to see that this construction produces
precisely those solutions associated to the subgrassmannian
$Gr_1\subset Gr$ \bibref[SW,W].

Moreover, the ring $A_C=\{p\in\C[z]|c_i\circ p\in C\ 1\leq i \leq n\}$
is necessarily non-trivial (i.e.\ contains non-constant polynomials) and the operator $L_p=p(\L)$ is an {\it
ordinary\/} differential operator for every $p\in A_C$ and satisfies
\begin{equation}
L_p\psi_C(x,z)=p(z) \psi_C(x,z).\label{eigenx}
\end{equation}
The subject of this paper is the existence of {\it additional\/} eigenvalue
equations satisfied by $\psi_C(x,z)$.  In particular, we wish to
consider the question of whether there exists an
operator $\hatLambda $ acting on functions of the variable $z$ such that
\begin{equation}
\hatLambda \psi_C(x,z)=\pi(x)\psi_C(x,z)\label{eigenz}
\end{equation}
where $\pi(x)$ is a non-constant function of $x$.
For example, the following theorem is due to G. Wilson in \bibref[W]:
\statetheorem[Th:wilson]{In addition to \eqref{eigenx} the wave function $\psi_C(x,z)$ is also an
eigenfunction for a ring of ordinary differential operators in $z$
with eigenvalues depending polynomially on $x$ if and only if $C$ has a
basis of distributions each of which is supported only at one point.}
In other words, for this special class of KP solutions for which the
coefficients of $\L$ are rational functions of $x$, the wave
function $\psi_C$ satisfies an additional eigenvalue equation of the form
\eqref{eigenz}
where $\hatLambda$ is an ordinary differential operator in $z$ and
$\pi(x)$ a non-constant polynomial in $x$.\footnote{Moreover, he
demonstrated that up to conjugation or change of variables, the
operators $L_p$ found in this way are the only bispectral operators
which commute with differential operators of relatively prime order,
but this fact will not play an important role in the present note.}
Together \eqref{eigenx} and \eqref{eigenz} are an example of {\it
bispectrality\/} \bibref[DG,G].  The bispectral property is already
known to be connected to other questions of physical significance such
as the time-band limiting problem in tomography \bibref[Gr1], Huygens'
principle of wave propagation \bibref[Yuri], quantum integrability
\bibref[HK,Vbis] and, especially in the case described above, the self
duality of the Calogero-Moser particle system \bibref[cmbis,W,W2].

It is known that the only subspaces $C$ for which the corresponding
wave function satisfies \eqref{eigenx} and \eqref{eigenz} with $L_p$ and
$\hatLambda $ ordinary differential operators in $x$ and $z$ respectively
are those described in Theorem~\ref{Th:wilson}.  However, suppose we
allow $\hatLambda$ to involve not only differentiation and multiplication
in $z$ but also {\it translation\/} in $z$ and  call this more {\it
general\/} situation t-bispectrality.\footnote{It should be noted that
the term ``bispectrality'' already applies to more general situations
than simply differential operators \bibref[G], but in the case of the
KP hierarchy I believe only differential bispectrality has thus far
been considered.}  It will be shown below that there are more KP
solutions which are bispectral in this sense.  In particular, a class
of (non-rational) $n$-soliton solutions of the KP hierarchy will be
shown to be t-bispectral.

\subsection{Notation}


 Using the shorthand notation 
$\partial=\frac{\partial}{\partial x}$ any ordinary differential
operator in $x$ can be
written as
$$
L=\sum_{i=0}^N f_i(x) \partial^i\qquad (N\in\N).
$$
All ordinary differential operators considered in this note will have
only coefficients that are rational functions of $x$ and of functions
of the form $e^{\lambda x}$.
Similarly, we will write
$\partial_z=\frac{\partial}{\partial z}$ but will need to consider
only differential operators in $z$ with rational coefficients.

For any $\lambda\in\C$ let $\Trans_{\lambda}=e^{\lambda\partial_z}$ be the
translational operator acting on functions of $z$ as
$$
\Trans_{\lambda}(f(z))=f(z+\lambda).
$$
Then consider the ring of translational-differential operators
$\TODOs$ generated by these translational operators and ordinary
differential operators in $z$.  Any translational-differential operator $\hat
T\in\TODOs$  can be
written as
$$
\genT =\sum_{i=1}^N p_i(z,\partial_z) \Trans_{\lambda_i}
$$
where $p_i$ are ordinary differential operators in $z$ with rational
coefficients  and $N\in\N$.  Note that the ring of ordinary
differential operators in $z$ with rational coefficients is
simply the subring of $\TODOs$ of all elements which can be written as
$p \Trans_{0}$ for a differential operator $p$.

\section{Translational Bispectrality of KP Solutions}

Let us say that a finite dimensional
subspace $C\subset\D$ is {\it t-bispectral\/} if there exists a
translational-differential operator  $\hatLambda \in \TODOs$ satisfying
equation \eqref{eigenz} for the corresponding KP wave function
$\psi_C(x,z)$.
By Theorem~\ref{Th:wilson} and the fact that the ring of rational coefficient
ordinary differential operators in $z$ is contained in $\TODOs$, we
know that $C$ is t-bispectral\footnote{...and that the ring $A_C$ is bispectral
in the sense of \bibref[W]} if it has a basis of point supported
distributions.  In order to prove that there are {\it other\/}
t-bispectral subspaces $C$, this section will determine a necessary
and sufficient
condition for t-bispectrality.

It will be convenient for us to refer to the following easily verified facts:

\statelemma[lem:zeroT]{A translational-differential operator
$\genT \in\TODOs$ satisfies $\genT (e^{xz})\equiv0$ if and only if $T\equiv 0$ is
the zero operator.}

\statelemma[lem:constcoef]{There exists a translational-differential operator
$\genT \in\TODOs$ satisfying $\genT (e^{xz})=\pi(x)e^{xz}$ for some
$z$-independent function $\pi(x)$ if and only if $\genT$ is a constant
coefficient operator and
\begin{equation}
\pi(x)=\sum_{i=1}^N p_i(x) e^{\lambda_ix}\qquad p_i\in\C[x],\ 
\lambda_i\in\C.\label{eqn:piform}
\end{equation}
}

\begin{lemma}[lem:twogammas]{Two translational-differential
operators $\hatLambda_1,\hatLambda_2\in\TODOs$ satisfy
\begin{equation}\hatLambda_1(e^{xz})=\pi(x)\hatLambda_2(e^{xz})
\label{eqn:justpi}
\end{equation}
 if and only if there
exist {\it constant coefficient\/} operators
$\hatGamma_1,\hatGamma_2\in\TODOs$ such that
\begin{equation}
\hatLambda_1\circ\hatGamma_1=\hat\Lambda_2\circ\hatGamma_2.
\label{eqn:twogammas}
\end{equation}}
First suppose that \eqref{eqn:justpi} is satisfied.
Then, since
$$
\pi(x)=\frac{\hatGamma_2(e^{xz})}{\hatGamma_1(e^{xz})}
$$
it is necessarily of the form $\pi(x)=\frac{\pi_2(x)}{\pi_1(x)}$ where
$\pi_i(x)$ are both functions of the form \eqref{eqn:piform}.  Then
let $\hatGamma_i$ be the constant coefficient
translational-differential operator such that
$\pi_i(x)=e^{-xz}\hatGamma_i(e^{xz})$.  Then, the operator
$\genT_0=\hatLambda_1\circ\hatGamma_1-\hatLambda_2\circ\hatGamma_2$
satisfies
$$
\genT_0(e^{xz})=\pi_1(x)\frac{\pi_2(x)}{\pi_1(x)}\hatLambda_2(e^{xz})-\pi_2(x)\hatLambda_2(e^{xz})=0$$
and so \eqref{eqn:twogammas} follow from Lemma~\ref{lem:zeroT}.

Conversely, it is clear that if \eqref{eqn:twogammas} is satisfied
then 
\begin{eqnarray*}
\hatLambda_1(e^{xz}) &=&
\frac{1}{\pi_1(x)}\hatLambda_1\circ\hatGamma_1(e^{xz})\\
 &=& \frac{1}{\pi_1(x)}\hatLambda_2\circ\hatGamma_2(e^{xz})\\
 &=& \frac{\pi_2(x)}{\pi_1(x)}\hatLambda_2(e^{xz})\\
 &=& \pi(x)\hatLambda_2(e^{xz}).
\end{eqnarray*}
\end{lemma}

An important object in much of the literature on integrable systems is
the ``tau function''.  The tau function of the KP solution associated
to $C$ can be written easily in terms of the basis $\{c_i\}$.  In
particular, define (cf.\ \bibref[W])
$$
\tau_C(x)=\textup{Wr}\left(c_1(e^{xz}),c_2(e^{xz}),\ldots,c_n(e^{xz})\right)
$$
to be the Wronskian determinant of the functions $c_i(e^{xz})$.
Similarly, there is a Wronskian formula for the coefficients of the
operator $K_C$ since its action on an arbitrary function $f(x)$ is
given as:
\begin{equation}
K_C(f(x))=\frac{1}{\tau_C(x)} \textup{Wr}\left(c_1(e^{xz}),c_2(e^{xz}),\ldots,c_n(e^{xz}),f(x)\right).\label{eqn:wrformK}
\end{equation}
Then
the coefficients of the differential operator $\tau_C(x)K_C(x,\partial)$
are all polynomials in $x$ and functions of the form $e^{\lambda x}$.
Consequently we have:

\statelemma[lem:hatT]{For any choice of $C$ there exists a translational-differential operator $
\hatTC\in\TODOs$ such that
\begin{equation}
\hatTC (e^{xz})=\tau_C(x)\psi_C(x)\label{eqn:hatTC}
\end{equation}}

Using these lemmas, it is possible to prove the following statement
which characterizes those subspaces $C$ which can be t-bispectral.

\begin{theorem}[thm:iff]{The operator $\hatLambda$ satisfies
\eqref{eigenz} if and only if there exist {\it constant coefficient\/}
translational-differential operators
$\hatGamma_1,\hatGamma_2\in\TODOs$ such that
\begin{equation}
\hatLambda\circ\hatTC\circ\Gamma_1=\hatTC\circ\Gamma_2.\label{eqn:notquitebdt}
\end{equation}}
Supposing that \eqref{eqn:notquitebdt} holds, it follows from 
Lemmas~\ref{lem:twogammas} and \ref{lem:hatT} that
$$
\hatLambda\psi_C(x,z)=\frac{1}{\tau_C(x)}\hatLambda\circ\hatTC(e^{xz})=\frac{\pi(x)}{\tau_C(x)}\hatTC(e^{xz})=\pi(x)\psi_C(x,z).
$$
Conversely, if \eqref{eigenz} is satisfied then
$$
\hatLambda\circ\hatTC(e^{xz})=
\tau_C(x)\hatLambda\psi_C(x,z)=
\tau_C(x)\pi(x)\psi_C(x,z)
=\pi(x)\hatTC(e^{xz})
$$
and then \eqref{eqn:notquitebdt} follows from Lemma~\ref{lem:twogammas}.
\end{theorem}

Of course, since it is already known that any subspace $C$ with a
basis of point supported distributions is bispectral,
Theorem~\ref{thm:iff} holds in those cases.  Moreover, in all of those
cases one is able to take $\hatGamma_1$ in \eqref{eqn:notquitebdt} to
be the identity operator.  So, this special case is clearly useful:
\statecorollary[cor:bdt]{The subspace $C\subset\D$ is t-bispectral if
there exist operators $\hatLambda,\hatGamma\in\TODOs$ such that
\begin{equation}
\hatLambda\circ\hatTC=\hatTC\circ\hatGamma\label{eqn:bdt}
\end{equation}
and such that $\hatGamma$ is a constant coefficient operator.}

In the context of bispectral ordinary differential operators, equation
\eqref{eqn:bdt} describes a {\it bispectral Darboux transformation\/}
\bibref[BHYpla,BHYcommrings,KR,bispdarb] of the operator $\hatGamma$ into the
operator $\hatLambda$.  The next section will demonstrate that the
same procedure may be used to construct examples of t-bispectral
subspaces corresponding to non-rational KP solutions.

\section{Non-rational t-Bispectral KP Solutions}

Let $$c=\sum_{k=1}^{N}
\alpha_k \Delta(\mu_k,n_k)\in\D\qquad\alpha_k\in\C$$ be an arbitrary finitely supported
distribution and pick any polynomial $q(z)\in \C[z]$.
We write $n=\deg q$ and label its roots and multiplicities by
$$
q(z)=\prod_{i=1}^{m}(z-\lambda_i)^{m_i}.\qquad \sum m_i=n.
$$
Then consider the $n$ dimensional subspace $C$ spanned by the distributions
$$c_{ij}=\sum_{k=1}^N \alpha_k
\Delta(\mu_k+\lambda_i,n_k+j-1)\qquad 1\leq i\leq m,\ 1\leq j \leq m_i.$$
The main result of this note is the
following:
\statetheorem[thm:main]{Every $C\subset\D$ determined from $c\in\D$
and $q\in\C[z]$ as described above is t-bispectral.  That is, there exists a translational-differential
operator $\hatLambda \in\TODOs$ in the variable $z$ such that the wave function $\psi_C$
satisfies \eqref{eigenz}.}

In the case that $c$ has support at a single point,
this merely reproduces the known result that there exists an ordinary
differential operator in $z$ having $\psi_C$ as an eigenfunction.
However, for general $c$ the other operator cannot be simply a
differential operator.  In particular, in the case that $c$ is chosen
to be of the form $c=\Delta(\lambda_1,0)+\hatGamma \Delta(\lambda_2,0)$
($\lambda_1\not=\lambda_2$) and $q$ is a polynomial with distinct
roots, then the corresponding solution is a pure $n$-soliton solution
of the KP hierarchy.

\subsection{Proof of Theorem~\ref{thm:main}}

The following lemma demonstrates
 that $K_C(x,\partial)$ can be written in a very
simple form for this particular choice of $C$ in terms of the
polynomial $q$ and the function $\phi(x):=c(e^{xz})$:
\begin{lemma}[lem:simpK]{For $C$, $q$ and $\phi$ as above, the dressing
operator $K_C$ can be written as
$$
K_C=\phi(x)q(\partial)\frac{1}{\phi(x)}.
$$}
This follows from the fact that $K$ is the unique monic operator of
order $n$ which annihilates all of the functions $c_{ij}(e^{xz})$.  Using
the easily derived formula
$$\frac{c_{ij}(e^{xz})}{\phi(x)}=x^{j-1}e^{x\lambda_i}$$
it follows that 
\begin{eqnarray*}
\left(\phi(x)q(\partial)\frac{1}{\phi(x)}\right)c_{ij}(e^{xz})
 &=& \phi(x)q(\partial)x^{j-1}e^{x\lambda_i}\\
 &=& \phi(x)\hat
q(\partial)(\partial-\lambda_i)^{m_i}x^{j-1}e^{x\lambda_i}\\
 &=& 0
\end{eqnarray*}
since $j-1<m_i$.
Clearly the operator applied above has the correct order
and leading coefficient and so it must be $K_C$.\end{lemma}


Note that the translational-differential operator
$$
\TC=\frac{1}{q(z)}\sum_{j=1}^N\alpha_j\partial_z^{n_j} \Trans_{\mu_j}
$$
satisfies 
\begin{equation}
\TC(e^{xz})=\frac{\phi(x)}{q(z)}e^{xz}. \label{eqn:TC}
\end{equation}
Finally, using $\TC$ along with $\hatTC$ from Lemma~\ref{lem:hatT} we
are able to construct operators $\hatGamma,\hatLambda\in\TODOs$ satisfying
\eqref{eqn:bdt}. 

\begin{lemma}{$\hatGamma=\TC\circ \hatTC$ is a constant coefficient
translational-differential operator.}
\begin{eqnarray*}
\hatGamma(e^{xz}) &=& \TC\circ \hatTC e^{xz}\\
 &=& \TC(\tau_C(x)\psi_C)\\
 &=& \tau_C(x)\TC K_C e^{xz}\\
 &=& \tau_C(x) \phi(x)\circ q(\partial)\circ \frac{1}{\phi(x)}
\TC(e^{xz})\\
 &=& \tau_C(x) \phi(x)\circ q(\partial)\circ \frac{1}{\phi(x)}
\frac{\phi(x)}{q(z)}(e^{xz})\\
 &=& \tau_C(x)\phi(x) e^{xz}.
\end{eqnarray*}
Then, by Lemma~\ref{lem:constcoef} we see that $\hatGamma$ must be a
constant coefficient operator.
\end{lemma}

Letting $\hatLambda =\hatTC\circ \TC\in \TODOs$ it is clear that $$
\hatTC\circ \hatGamma = \hatTC \circ \TC\circ\hatTC= \hatLambda \circ
\hatTC.
$$ 
Thus, Theorem~\ref{thm:main} follows from Corollary~\ref{cor:bdt} and
we see that 
$$
\hatLambda\psi_C(x,z)=\tau_C(x)\phi(x)\psi_C(x,z).
$$

\section{Examples}
If we choose $c=\Delta(1,0)$ (corresponding to the
stationary rational KdV solution $u=-2/x^2$) and $q=z(z-1)$ then
$C$ is spanned by $c_1=c$ and $c_2=\Delta(1,1)$ (a ``two-particle''
Calogero-Moser type solution).  Now $\phi(x)=x$ and then
$$
\psi_C(x,z)=(1+\frac{2+x-(2x+x^2)z}{x^2z^2})e^{xz}.
$$
Obviously if $\TC=\frac{1}{q(z)}\partial_z\Trans_0$ then $\TC e^{xz}=\frac{x}{q(z)}e^{xz}$ and if
 $\hatTC=\frac{1}{z^2}((z^2-z)\partial_z^2+(1-2z)\partial_z+2)\Trans_0$ then
$\hatTC(e^{xz})=x^2\psi_C(x,z)$.  Then it turns out that the
operator $\hatLambda $ given by 
$$
\hatLambda =\hatTC\circ \TC 
$$
is simply the ordinary differential operator
$$
\hatLambda =\partial_z^3+\frac{3}{z-z^2}\partial_z^2-\frac{6z^2-12z+3}{z^3(z-1)^2}\partial
+ \frac{12z-6}{z^2(z-1)^2}$$
which satisfies $\hatLambda \psi_C(x,z)=x^3\psi_C(x,z)$ (as we would
expect from earlier results on bispectrality.)

However, if we had chosen instead $c=\Delta(0,1)+\Delta(0,-1)$, then the KP solution
corresponding to the space spanned by $c$ alone is a standard
1-soliton solution of the KdV hierarchy.  Letting $q=z(z-1)$ again we consider 
the space $C$ spanned by $c_1=c$ and
$c_2=\Delta(0,2)+\Delta(0,0)$  which
corresponds to a special case of the KP 2-soliton solution.

In this case we find that $\phi(x)=e^x+e^{-x}$ and so
$$
\psi_C(x,z)=(1-\frac{6+(3z-2)e^{2x}+2z-ze^{-2x}}{\phi^2(x)z^2})e^{xz}.
$$
One may easily check that
$
\TC=\frac{1}{q(z)}\Trans_1+\frac{1}{q(z)}\Trans_{-1}
$
satisfies \eqref{eqn:TC}.  Moreover, 
$$
\hatTC=(1-\frac3z-\frac2{z^2})\Trans_2-(1+\frac1z)\Trans_{-2}+(2-\frac2z-\frac6{z^2})\Trans_0
$$
satisfies \eqref{eqn:hatTC}.

Finally, one finds that 
\begin{eqnarray*}
\hatLambda  &=& (1+\frac1z)\Trans_{-3}+(3-\frac{6}{z^2}-\frac1z) \Trans_{-1}
\\
&& + (3-\frac{4}{z^2}-\frac5z)\Trans_0+(1+\frac{2}{z^2}-\frac3z)\Trans_3
\end{eqnarray*}
satisfies $\hatLambda \psi_C(x,z)=\phi^3(x)\psi_C(x,z)$.

\section{Conclusions}

In addition to being a generalization of the results of \bibref[DG,W]
on bispectral ordinary differential operators, the present note may be
seen as a generalization of \bibref[Reach] in which wave functions of
$n$-soliton solutions of the KdV equation are shown to satisfy
difference equations in the spectral parameter.  (In fact, the methods
of that paper are quite similar in many ways to the methods used here.)

As in \bibref[DG,W], the equations \eqref{eigenx} and \eqref{eigenz}
lead to the well known ``ad'' relations associated to bispectral
pairs.
That is, defining the ordinary differential operator $A_m$ in $x$ and
the translational-differential operator $\hat A_m$ in $z$ by 
$$A_m=\textup{ad}_{L_p}^m(\pi(x))
\qquad
\hat A_m=(-1)^m\textup{ad}_{p(z)}^m(\hatLambda)
$$
one finds that $A_m\psi_C(x,z)=\hat A_m\psi_C(x,z)$.  Similarly, if
$$B_m=\textup{ad}_{\pi(x)}^m(L_p)
\qquad
\hat B_m=(-1)^m\textup{ad}_{\hatLambda}^m(p(z))
$$
then $B_m\psi_C(x,z)=\hat B_m\psi_C(x,z)$.  Since the order of the
operator $B_m$ is at least one less than the order of the operator
$B_{m-1}$, the familiar result that $B_m\equiv0$ and $\hat B_m\equiv0$
for $m>\ord L_p$ holds, which is clearly a strong restriction on
the operator $\hatLambda$.  However, unlike the case of bispectral
ordinary differential operators, one cannot conclude that $A_m\equiv0$
for sufficiently large $m$ since the order of $\hat A_m$ may not be
reduced by increasing $m$. 


Finally, although the results of the preceding section demonstrate
that the generalization from ordinary bispectrality to t-bispectrality
is not a trivial one, they have not completely addressed the problem
posed above.  Namely, it remains to be determined whether there are
other t-bispectral subspaces $C$ and more significantly whether this
non-local form of bispectrality for KP solutions has any dynamical
significance.  In particular, just as Wilson's bispectral involution
is a manifestation of the self-duality of the Calogero-Moser system
\bibref[cmbis,W,W2], it seems reasonable to conjecture that the form
of bispectrality presented above should also be related to the
dualities of classical integrable particle systems.

\par\medskip\par\noindent{\bf Acknowledgements:}  I am grateful for support from and
conversations with John Harnad and the collaboration of Yuri Berest.
The observation that the dressing operator $K_C$ can have the special
form used above is a by-product of joint work with Y.B. on higher
dimensional Darboux transformations.

\end{document}